\documentclass[a4paper,fleqn,usenatbib]{mnras}

\usepackage{newtxtext,newtxmath}

\usepackage[T1]{fontenc}
\usepackage{ae,aecompl}
\usepackage{graphicx}	

\title[Radius Valley]{Examining the Radius Valley: a Machine Learning Approach}

\author[M. G. MacDonald]{Mariah G. MacDonald,$^{1,2}$\thanks{E-mail: mmacdonald@psu.edu}
\\
$^{1}$Department of Astronomy \& Astrophysics, The Pennsylvania State University, 525 Davey Lab, State College, PA, 16802, USA\\
$^{2}$Center for Exoplanets and Habitable Worlds, The Pennsylvania State University, 525 Davey Lab, State College, PA, 16802, USA
}

\date{Accepted XXX. Received YYY; in original form ZZZ}
\pubyear{2019}

\begin{document}
\label{firstpage}
\pagerange{\pageref{firstpage}--\pageref{lastpage}}
\maketitle

\begin{abstract}
 The ''radius valley" is a relative dearth of planets between two potential populations of exoplanets, super-Earths and mini-Neptunes. This feature appears in examining the distribution of planetary radii, but has only ever been characterized on small samples. The valley could be a result of photoevaporation, which has been predicted in numerous theoretical models, or a result of other processes. Here, we investigate the relationship between planetary radius and orbital period through 2-dimensional kernel density estimator and various clustering methods, using all known super-Earths ($R<4.0R_E$). With our larger sample, we confirm the radius valley and characterize it as a power law. Using a variety of methods, we find a range of slopes that are consistent with each other and distinctly negative. We average over these results and find the slope to be  $m=-0.319^{+0.088}_{-0.116}$. We repeat our analysis on samples from previous studies. For all methods we use, the resulting line has a negative slope, which is consistent with models of photoevaporation and core-powered mass loss but inconsistent with planets forming in a gas-poor disk.
\end{abstract}


\section{Introduction}

With the launch of \emph{Kepler} in 2009 and the following years of analysis, the astronomical community witnessed the number of known exoplanets skyrocket from a couple hundred to a couple thousand \citep{Lissauer2011,Lissauer2014,Fabrycky2014}. Newfound discoveries such as compact systems \citep{Lissauer2011,Muirhead2015}, resonant chains \citep{Lithwick2012,Fabrycky2014}, and low density planets \citep[e.g.,][]{Welsh2015,Weiss2014,Ford2017} led to numerous studies of the exoplanet population as a whole, as well as countless studies focused on just one or a few systems. Many of these studies aimed to better characterize exoplanets and estimate the planet parameters to gain insight into planet formation.

Several theoretical models predict that the size of planets can be strongly influenced by the host stars. In particular, formation models find that atmospheric erosion, which mostly affects planets near their star, will result in a so-called ''photoevaporation valley,'' a gap in the distribution of planetary radii around $2R_E$ \citep{Owen2013,Jin2014,Lopez2014,Chen2016,Lopez2016,Owen2017}. This valley is established by the boundary between planets that are massive enough to keep their gas envelopes and planets whose atmospheres are stripped away. The specific slope of the valley depends on various factors, such as the composition of the planets and the physics of the evaporation \citep[e.g.,][]{Owen2017}, but the overall sign of the slope is negative. In particular, the separation appears linear in a log-log plot of planetary radius vs. orbital period:

\begin{equation}
    \textrm{log}~R_p = m\cdot \textrm{log}~P + b
\end{equation}

\noindent where $P$ is the orbital period, $R_p$ is the planetary radius, $m$ is the slope in log-log space, and $b$ is the y-intercept in log-log space. In linear space, this translates to a power law $R_p = e^b\cdot P^m$.

Actually observing this valley can be challenging, as uncertainties in stellar parameters can obfuscate its presence and render characterization of the valley, to aid in understanding planet formation, quite difficult. \citet{Fulton2017} first observed the gap in distribution of planetary radii using precise radius measurements from the California-Kepler Survey and a sample of 2025 exoplanets. They note the gap as a dearth of planets with radii between 1.5 and 2.0 $R_E$ and suggest that this limit essentially separates super-Earths from mini-Neptunes, meaning that it determines whether a planet has a substantial H/He envelope. Using Gaia parallaxes, \citet{Fulton2018} released an updated catalog of $\sim$1000 planets with improved precision in the planet radius measurements ($\sim$5\%). They found that the mass of the stellar host and the planet's orbital distance are both key factors that sculpt the distribution of planetary radii, concluding that the gap does indeed have planets in it and is more a dearth of planets around $R_p\sim2R_E$. \citet{Berger2018} also released an updated catalog with stellar radii derived from a combination of Gaia DR2 parallaxes and the DR25 Kepler Stellar Properties Catalog. They confirm a gap in the distribution of radii with their updated values, noting that the gap is mostly limited to large incident fluxes ($>200 F_E$). These three studies did not attempt to constrain the valley as a function of orbital period.

Soon after \citet{Fulton2017} released their study, \citet{vaneylen2018} explored the location and shape of the radius valley using 117 planets with accurate parameters determined from asteroseismology. Using a linear model (in log-log space) and an MCMC algorithm, they first characterize the valley by fitting the absence of data, recovering a slope of $m=-0.10\pm0.03$. They then use support vector machines to determine the hyperplane of maximum separation between the planets above and below the valley, resulting in a slope of $m=-0.09^{+0.02}_{0.04}$. They conclude that the negative slope they measured is consistent with models of photoevaporation.

There are a few possible explanations for this radius valley aside from mass loss by photoevaporation, including  sculpting by giant impacts \citep[e.g.,][]{Liu2015,Schlichting2015,Inamdar2016}, core-powered mass loss \citep[e.g.,][]{Ginzburg2017,Gupta2018}, and planets forming late in a gas-poor disk \citep[e.g.,][]{Lee2014,Lee2016}. It is unclear whether impacts alone can cause a gap in the distribution of planetary radii since impact erosion is a highly stochastic process, but atmospheric heating caused by an impact can cause the envelope to expand, which would make it more susceptible to photoevaporation. Core-powered mass loss results in a negative slope in the valley and can sufficiently explain the gap in the distribution of planetary radii \citep{Ginzburg2018,Gupta2018}. \citet{Lopez2016} explore the idea that super-Earths near their stars are actually a separate population of planets that developed late in a gas-poor disk and never accreted significant envelopes, instead of accreting atmospheres that are then later stripped away. They found that the transition radius between this population and other planets depends on orbital period $R\propto P^{m}$, where $m$ is between 0.07 and 0.10. Therefore, forming in a gas-poor disk would result in the slope of a radius valley that is positive.

To date, there has been no observational study that looks to characterize the radius valley, in both radius and orbital period, without suffering from a small sample size. Here, we aim to characterize the valley by measuring its slope and to identify the dominant mechanism that is responsible for sculpting the radii of exoplanets. We investigate the existence and slope of the radius valley using all known exoplanets with orbital period and planetary radius measurements of $P<50,~R_p<4.0R_E$. 

The outline of this paper is as follows: in Section~\ref{sec: data}, we describe the data used in this study. In Section~\ref{sec: methods}, we describe and explain the theories behind the methods and tests used to examine the radius valley. Following this, we show and discuss the results from our analysis in Section~\ref{sec: results} and Section~\ref{sec:disc} before concluding in Section~\ref{sec: conc}.


\section{Data}\label{sec: data}

We use all known exoplanets in our analyses, pulling data from the exoplanetarchive\footnote{exoplanetarchive.ipac.caltech.edu, data pulled on April 5, 2019}. Although exoplanetarchive keeps track of all published parameters for planets, we use only the most recent parameters for the study presented in this paper. We include planets from all detection methods, although \textit{Kepler} contributed the most data, resulting in 3933 known exoplanets. We only use planets with published values and uncertainties for radius and orbital period and limit our sample to $P<50$ days and $R_p<4.0R_E$, resulting in 2066 planets. The Solar System was not included in this study. 

We perform our analysis on other data sets that have been used to study this phenomenon in the past, particularly those of  \citet{Fulton2017}, \citet{vaneylen2018}, and \citet{Fulton2018}. We refer to these data sets as ''F17,'' ''V18,'' and ''F18'' from here on. We present information about each data set, including number of planets and average uncertainty in planetary radius, in Table~\ref{tab:data}.

\begin{table}
    \centering
    \begin{tabular}{lccc}
    \hline
    \hline
Data set &	Sample Size	&	$<R_{\sigma}>$ ($R_E$)	& Ref	\\			
\hline	
All & 2066 & 0.252 & -- \\
F17 & 766 & 0.19 & \citet{Fulton2017} \\
V18 & 81 & 0.0615 & \citet{vaneylen2018} \\
F18 & 1334 & 0.089 & \citet{Fulton2018} \\

\hline																										
	\end{tabular}
    \caption{The number of planets and average (median) uncertainty in planetary radius for planets in each of the data sets we use. For data set ''all,'' we pull all exoplanets from the exoplanet archive (April 5, 2019). We only use planets with published uncertainties in both planetary radius and in orbital period. We also limit all samples to super-Earths ($R<4.0R_E$) with orbital periods $P<50$ days. }
    \label{tab:data}
\end{table}


\section{Methods and Theory }\label{sec: methods}
We analyze the exoplanetary population using 2-D kernel density estimator, hierarchical clustering, $K$-means clustering, $K$-medians clustering, and density-based clustering. The methods used here and the theories behind them are briefly described below. We perform all of our analysis in the R language \citep{Rstats}. 

\subsection{Kernel Density Estimator}
A kernel density estimation is a nonparametric estimation of probability density functions. The kernel density estimate is defined by:

\begin{equation}
    \hat{f}(x;H) =n^{-1}\sum^n_{i=1}K_H(x-X_i)
\end{equation}
\noindent where $x=(x_1,x_2)^T$ and $X_i = (X_{i1},X_{i2})^T$ for $i$=1,2,...,n a bivariate random sample drawn from a density $f$, $K(x)$ is the kernel which is a symmetric probability density function, and H is the bandwidth matrix which is symmetric and positive-definite. Although the choice of the kernel is not very important, in this analysis we use an axis-aligned bivariate normal kernel, evaluated on a square grid.  A KDE is dependent on the distribution of data given to it, so our results will be sensitive to incompleteness and effects from biases. We also must choose a bandwidth for both orbital period and planetary radius; the results of the KDE are not very dependent on the bandwidths, but bandwidths that are too large will result in a uni-modal distribution and bandwidths that are too small will result in multi-modal distributions. We choose bandwidths that lead to bi-modal distributions so that we are able to characterize a line between the modes. In our analysis, we use the function \textit{kde2d} from CRAN package \verb|MASS| \citep{MASS}. For a more in-depth description of KDEs and their applications, we encourage the reader to explore \citet{silverman2018}.

\subsection{Hierarchical Clustering}
Hierarchical clustering is a method of cluster analysis that builds a hierarchy of clusters. Originally, each object receives its own cluster; the algorithm computes the distance between clusters via the Lance-Williams dissimilarity formula and combines the two most similar clusters. The Lance-Williams algorithms are a family of agglomerative clustering algorithms which are represented by a recursive formula. After two clusters are merged, the updated cluster distance is computed recursively:

\begin{equation}
    d_{(ij)k} = \alpha_id_{ik} + \alpha_j d_{jk} + \beta d_{ij} + \gamma|d_{ik}-d_{jk}|
\end{equation}

\noindent where $i$ and $j$ indicate the clusters being merged, $d$ is the cluster distance and $\alpha_i$, $\alpha_j$, $\beta$, and $\gamma$ are parameters. We use complete linkage method to find similar clusters, where the distance between two clusters is the distance between the two elements (one in each cluster) that are farthest away from each other. This process continues until all objects have been sorted into one cluster, and a decision tree has been formed. Simply put:

1. Assign every point to its own cluster.

2. Compute the distance between each cluster.

3. Merge the two closest clusters.

4. Update the distances between the new cluster and 

the original clusters.

5. Repeat steps 3 and 4 until only a single cluster 

remains.

\noindent We cut the hierarchical tree at two branches so that we have two populations (super-Earths and mini-Neptunes). One caveat to this method is we cannot control along which access the resulting two clusters are separated. In our analysis, we use the function \textit{hclust} from CRAN package \verb|MASS| \citep{MASS}. We refer the reader to \citet{Johnson1967} for a discussion on various hierarchical clustering schemes. 

\subsection{K-means Clustering}\label{sec:kmeans}
Originally from signal processing, $K$-means clustering is a method of vector quantization. We use the original algorithm by \citet{Hartigan1979} which aims to separate the data into $K$ number of groups such that the sum of squares from all points to the cluster centers is minimized. Mathematically, the algorithm aims to find:

\begin{equation}
    \substack{\textrm{arg min } \\ \bar{S}} \sum^K_{i=1} \sum_{\bar{x}\in S_i} ||  \bar{x}-\mu_i  || ^2 = \substack{\textrm{arg min} \\ \bar{S}} \sum^K_{i=1} | S_i | \textrm{ Var } S_i
\end{equation}

\noindent where $\bar{x}=\{x_1,x_2,...,x_n \}$ is a data set, $\bar{S}=\{S_1,S_2,...,S_k \}$ is a set of clusters, and $\mu_i$ is the mean of points in $S_i$. Simply put:

1. Select $K$ points as the initial centeroids.

2. Assign all points to the closest centeroid (mean point 

of cluster).

3. Recompute the centroid (mean) of each cluster.

4. Repeat steps 2 and 3 until the centroids no longer 

change.

\noindent We choose $K=2$ for two clusters. In our analysis, we use the function \textit{kmeans} from base R \citep{Rstats}. We refer the reader to \citet{macqueen1967} or \citet{Hartigan1979} for a more thorough description of the algorithm.

\subsection{K-medians Clustering}
$K$-medians clustering is another method of vector quantization. We use the same algorithm as presented above in Section~\ref{sec:kmeans}. The only difference between the two methods is $\mu_i$ is now the median of the points in each cluster $S_i$, so the centeroid is the median point of the cluster. We choose $K=2$ for two clusters.  In our analysis, we use the function \textit{kGmedian} from CRAN package \verb|Gmedian| \citep{Gmedian}. We encourage the reader to explore \citet{mulvey1979} for a detailed description and analysis of this method.

\subsection{Density-based Clustering}
Density Based Spatial Clustering of Applications with Noise (DBSCAN) is a clustering technique that was designed to discover the clusters and the noise in a spatial database \citep{ester1996}. The aim of the algorithm is for each point of a cluster in the ''neighborhood'' of a given radius $\epsilon$ to contain at least a minimum number of points. In other words, the density in the ''neighborhood'' has to exceed some threshold. The algorithm works as:

1. Start with an arbitrary point $p$ in the data and 

retrieve all points that are ''density-reachable,'' 

i.e. the other point belongs in the ''neighborhood'' 

of the initially selected point $p$ and the 

''neighborhood'' contains the minimum number of  

points.

2. Repeat step 1 until some points are density-reachable.

3. Calculate the distances between the resulting 

clusters. 

4. Merge clusters that are at least as dense as the 

thinnest cluster if the two clusters are closer than 

the given radius $\epsilon$.

5. Repeat steps 3 and 4 until no more mergers occur.

\noindent The choice of distance function that is used will determine the shape of the ''neighborhood,'' but the resulting clusters can have any shape \citep{ester1996}. We use the Manhattan distance in 2D space which results in a rectangular neighborhood.  We randomly draw our $\epsilon$ and minimum number of points from uniform distributions such that the clustering results in two populations. DBSCAN specifically will account for noise in the data, where noise is simply the data points that do not belong to any clusters. This method is therefore less sensitive to outliers than other methods \citep{ester1996,campello2013}. It also means that the clusters are smaller (spatially, and contain fewer points) than those from the other methods which force all points into a cluster. In our analysis, we use the function \textit{dbscan} from CRAN packages \verb|dbscan| \citep{dbscan}. We refer the reader to \citet{ester1996} for the full mathematical definition and implementation of the algorithm.

\subsection{A Comparison and Caveats of Clustering Techniques}
We use a variety techniques so that our resulting slope is not dependent on the technique. However, many computer scientists and statisticians have been studying and comparing these methods for decades. The general consensus is that a variety of clustering techniques should be studied for each new data set, given that different methods are sensitive to different pit-falls \citep[e.g.,][]{ester1996,chen2002,mingoti2006,sharma2012}.

In general, hierarchical clustering or density-based clustering are considered to be superior when the number of clusters is unconstrained or poorly constrained. Partitioning methods such as $K$-means or $K$-medians, however, can be much faster and less computationally intensive \citep[e.g.,][]{chen2002,sharma2012}.  For the problem presented here, we know that we want two populations (super-Earths and mini-Neptunes) so this superiority no longer applies. Instead, the hierarchical clustering introduces a bit of a problem in our slope determination: given that all of the data have been sorted into one cluster via a tree, and we simply cut this tree at two branches, the axis of separation between the two clusters can readily be along orbital period instead of planetary radius. Oftentimes, the two populations will be separated along $R_p\sim2R_E$, but sometimes the populations will be separated by $P\sim7$ days. This results in lines that are near-vertical with very negative slopes. 

\citet{chen2002} found that, in analyzing ES cell gene expression data, hierarchical clustering performed worse than $K$-means clustering. They attributed its poor performance to the algorithm's ''greediness'' since it is not possible to do any refinement or correction once two clusters have merged. Other studies have found similar results  \citep[e.g.,][]{mingoti2006}.

DBSCAN has been noted to be inefficient for high-dimensional data sets, but is often praised for its notion of noise and ability to find arbitrarily shaped clusters, including clusters that are completely surrounded by other clusters \citep{ester1996,sharma2012,campello2013}. It has also been found to be at least 100 times more efficient than CLARANS, a $K$-medoid algorithm similar to $K$-means and $K$-medians  \citep{ester1996}. 

Both $K$-means and $K$-medians clustering are typically faster than hierarchical clustering, especially for large data sets \citep{ester1996,sharma2012}, but do not work very well with non-globular clusters \citep{sharma2012}. $K$-means clustering is also affected by the presence of a large amount of outliers $\sim$40\% \citep{mingoti2006}. The accuracy of $K$-means and $K$-medians clustering is also quite dependent on the choice of initial centeroids \citep{milligan1980}. To help mitigate this dependence, our bootstrapping technique starts with different centeroids each time for both methods. Our clusters are also fairly globular in shape and do not contain so many outliers, so these pit-falls do not effect our results.


\section{Results}\label{sec: results}

Here we analyze the resulting slopes from our various methods and data sets. We report our bestfit slope and its uncertainty as the median and 16th and 84th percentiles from a combination of all of our bootstrap realizations from using all known exoplanets with period $P<50$ days and radius $R_p < 4.0 R_E$. We report the resulting slopes from our different data sets and methods in Table~\ref{tab:slopes}.

\subsection{Recovering a Line from Between the Two Populations}\label{sec:line}

We aim to recover the slope of the linear separation between the two populations to help determine the cause of this observed dichotomy. For the 2-D KDE, we perform a form of bootstrapping to determine the median and uncertainty in the slope. For each of our 200 bootstrap samples, we draw (with replacement) 2000 planets from the data set. We randomly select our bandwidths for orbital period and for planetary radius from U[2.0,15.0] days and U[0.2,0.4] $R_E$, respectively, for each realization. These bandwidths mark the limits for a bi-modal distribution.  For each realization, we first find the line between the two peaks in the full period-radius distribution, where period is in days and planetary radius is in $R_{\oplus}$. We then take the line that is perpendicular to this line to be our separator. 

For each of our clustering methods (hierarchical, $K$-means, $K$-medians, density-based), we fit a line separating the two categories using a linear discriminant analysis, a generalization of Fisher's linear discriminant \citep{Fisher1936} that aims to find a linear combination of features that separates two populations of objects. Although an in-depth derivation of this process is beyond the scope of this work, we refer the reader to \citet{mika1999} for an excellent review. We again perform a form of bootstrapping where we draw (with replacement) 2000 planets from the data set 200 times. For the density-based clustering, we choose a bandwidth $\epsilon$ from U[0.315,0.362] for each realization which results in two clusters. 

\subsection{Resulting slopes and intercepts}

\begin{figure}
    \centering
    \includegraphics[width=0.48\textwidth]{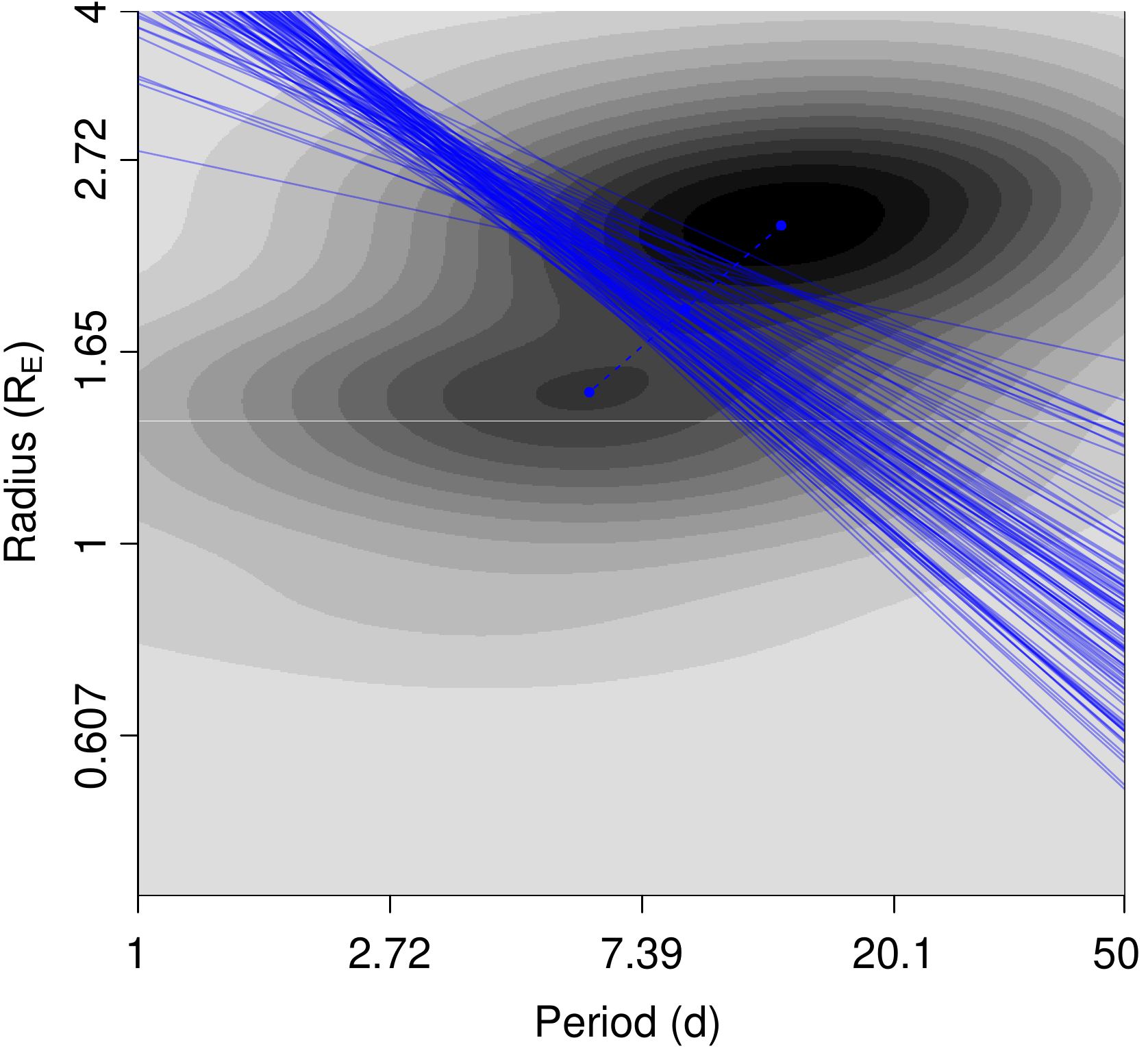}
    \caption{2D Kernel Density estimation of planetary radius versus orbital period for all known exoplanets with period $P<50$ days and radius $R_p < 4.0 R_E$. We did not use planets that did not have published values for period, radius, or uncertainty in radius. We performed a bootstrapping variant, varying the bandwidth for period between 2.0 and 15.0 days and the bandwidth for radius between 0.2 and 0.4 $R_E$, although the resulting slopes were not very dependent on the bandwidths. For each permutation, we also draw (with replacement) 1000 planets.  We account for uncertainties in radius by drawing the planets' radii from normal distributions where the means were the published values and the standard deviations were the published uncertainties. Here, we plot the resulting 200 slopes overtop a kernel density estimation with bandwidths of 13.5 days and 0.4$R_E$. The resulting slope estimation for ''all'' data was $m=-0.35	^{+0.15	}_{-0.14	} $, suggesting that the radius valley we see is likely an artifact of photo-evaporation or core-powered mass loss and not of planets forming in a gas-poor disk. }
    \label{fig:kde}
\end{figure}

We apply a two-dimensional kernel density estimator (2-D KDE) to the data with an axis-aligned bivariate normal kernel, evaluated on a square grid. We perform a bootstrapping variant, sampling our data and measuring the line between populations 200 times. For each sample, we draw with replacement 2000 planets from the data set (see Section~\ref{sec: data} for a description of each set). We plot the KDE with the resulting separators in Figure \ref{fig:kde}.  The resulting slope estimation for ''all'' data is $m=-0.35^{+0.15}_{-0.14} $ where our uncertainties mark the 16th and 84th percentiles. We present the slope estimates and uncertainties for all methods and data sets in Table~\ref{tab:slopes}.

\begin{figure}
    \centering
    \includegraphics[width=0.48\textwidth]{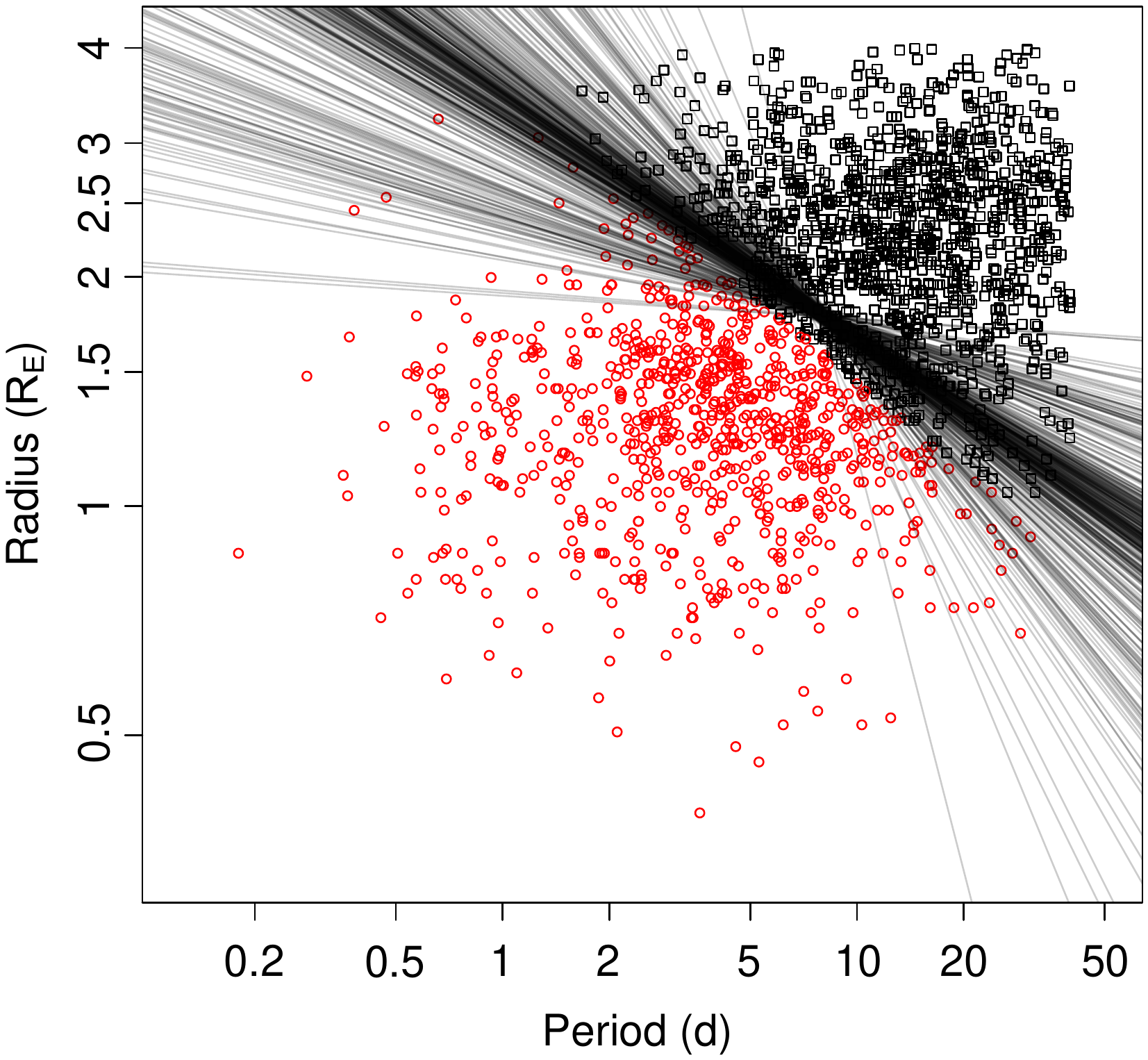}
    \caption{The result of $K$-means clustering, with lines defining the radius valley from all bootstrap realizations from all methods (2000 lines) for all known exoplanets with period $P<50$ days and radius $R_p < 4.0 R_E$. The radius valley does not appear to be much of a valley, since it is not a complete absence of planets in that region of parameter space, but  clustering does split the planets into two populations, usually about $R\sim2R_{\oplus}$. Different methods result in different slopes and precisions, but all methods and data sets are consistent with a negative slope, suggesting that the radius valley is likely a result of photoevaporation \citep[e.g.,][]{Owen2017} or core-powered mass loss \citep[e.g.,][]{Gupta2018} and not of planets forming in a gas-poor disk. }
    \label{fig:clust}
\end{figure}

We next cluster our data using a variety of methods described in Section~\ref{sec: methods}: hierarchical clustering, $K$-means clustering, $K$-medians clustering, and density-based clustering. For each of our clustering methods, we repeat the bootstraping variant as mentioned above, selecting 2000 planets with replacement from the data set. We then cluster the data into two clusters for the $K$-means and $K$-medians clustering. For the hierarchical clustering, we cut the tree at two branches since hierarchical clustering clusters the data into an entire tree of classification. This sometimes led to the clustering being solely period-based, resulting in very large and negative slopes. For each clustering method and each bootstrapped realization, we fit a line between the two clusters, as described above in Section~\ref{sec:line}, and take the median slope to be our bestfit and the 16th and 84th percentiles as our uncertainties in the slope. The resulting slope estimation between the two populations for ''all'' data is $m=-0.316	^{+0.077}_{-0.084} $ for all clustering methods with a resulting y-intercept of $b=1.26^{+0.28}_{-0.17}$. We show a realization of clustering with all fitted lines plotted on top in Figure~\ref{fig:clust}. 

We present the slope estimates and uncertainties for all methods and data sets in Table~\ref{tab:slopes}. We take our bestfit slope to be the result of ''all'' data set and all methods $m=-0.319^{+	0.088	}_{-	0.116	} $. All methods and data sets result in consistent and distinctly negative slopes, which are consistent with both models of photoevaporation \citep[e.g.,][]{Owen2017} and core-powered mass loss \citep[e.g.,][]{Gupta2018}, but inconsistent with planets forming late in a gas-poor disk.

\begin{table*}
    \centering
    \begin{tabular}{lcccccccc}
    \hline
    \hline
		
Data set	&	Num. points	&	Kmeans					&	Kmedians					&	DBSCAN					&	HIER					&	KDE					&	CLUST					&	ALL	M			\\
\hline																																													
All	&	2066	& $	-0.329	^{+	0.037	}_{-	0.039	} $ & $	-0.306	^{+	0.029	}_{-	0.031	} $ & $	-0.324	^{+	0.085	}_{-	0.087	} $ & $	-0.253	^{+	0.127	}_{-	0.307	} $ & $	-0.356	^{+	0.146	}_{-	0.144	} $ & $	-0.316	^{+	0.077	}_{-	0.084	} $ & 	\textbf{-0.319}	$^{+	0.088	}_{-	0.116	} $ \\
All, $\sigma_r$	&	1000	& $	-0.590	^{+	0.181	}_{-	0.331	} $ & $	-0.651	^{+	0.160	}_{-	0.282	} $ & $	-0.658	^{+	0.332	}_{-	0.587	} $ & $	-0.366	^{+	0.193	}_{-	0.636	} $ & $	-0.333	^{+	0.166	}_{-	0.185	} $ & $	-0.608	^{+	0.291	}_{-	0.371	} $ & $	-0.531	^{+	0.265	}_{-	0.401	} $ \\
All P$<25$	&	1855	& $	-0.398	^{+	0.042	}_{-	0.050	} $ & $	-0.366	^{+	0.036	}_{-	0.038	} $ & $	-0.530	^{+	0.120	}_{-	0.177	} $ & $	-0.221	^{+	0.104	}_{-	0.373	} $ & $	-0.403	^{+	0.090	}_{-	0.059	} $ & $	-0.393	^{+	0.104	}_{-	0.159	} $ & $	-0.397	^{+	0.096	}_{-	0.123	} $ \\
All P$<25$, $\sigma_r$	&	1000	& $	-0.793	^{+	0.304	}_{-	0.243	} $ & $	-0.745	^{+	0.191	}_{-	0.155	} $ & $	-0.725	^{+	0.198	}_{-	0.276	} $ & $	-0.603	^{+	0.325	}_{-	0.684	} $ & $	-0.340	^{+	0.074	}_{-	0.106	} $ & $	-0.738	^{+	0.281	}_{-	0.288	} $ & $	-0.640	^{+	0.293	}_{-	0.352	} $ \\
F17	&	766	& $	-0.525	^{+	0.059	}_{-	0.074	} $ & $	-0.450	^{+	0.042	}_{-	0.066	} $ & $	-0.480	^{+	0.197	}_{-	0.143	} $ & $	-0.783	^{+	0.378	}_{-	0.237	} $ & $	-0.420	^{+	0.019	}_{-	0.020	} $ & $	-0.508	^{+	0.116	}_{-	0.227	} $ & $	-0.474	^{+	0.076	}_{-	0.183	} $ \\
F17, $\sigma_r$	&	1000	& $	-0.379	^{+	0.058	}_{-	0.078	} $ & $	-0.348	^{+	0.032	}_{-	0.045	} $ & $	-0.146	^{+	0.056	}_{-	0.068	} $ & $	-0.236	^{+	0.172	}_{-	0.607	} $ & $	-0.378	^{+	0.023	}_{-	0.032	} $ & $	-0.326	^{+	0.202	}_{-	0.120	} $ & $	-0.348	^{+	0.205	}_{-	0.086	} $ \\
V18	&	81	& $	-0.561	^{+	0.144	}_{-	0.196	} $ & $	-0.553	^{+	0.025	}_{-	0.025	} $ & $	-0.369	^{+	0.066	}_{-	0.120	} $ & $	-0.272	^{+	0.018	}_{-	0.018	} $ & $	-0.315	^{+	0.026	}_{-	0.042	} $ & $	-0.476	^{+	0.204	}_{-	0.109	} $ & $	-0.376	^{+	0.100	}_{-	0.194	} $ \\
V18, $\sigma_r$	&	1000	& $	-0.565	^{+	0.039	}_{-	0.039	} $ & $	-0.570	^{+	0.031	}_{-	0.026	} $ & $	-0.362	^{+	0.061	}_{-	0.073	} $ & $	-0.226	^{+	0.021	}_{-	0.430	} $ & $	-0.328	^{+	0.013	}_{-	0.015	} $ & $	-0.535	^{+	0.292	}_{-	0.066	} $ & $	-0.433	^{+	0.141	}_{-	0.160	} $ \\
F18	&	1334	& $	-0.370	^{+	0.038	}_{-	0.058	} $ & $	-0.334	^{+	0.035	}_{-	0.030	} $ & $	-0.303	^{+	0.086	}_{-	0.101	} $ & $	-0.267	^{+	0.167	}_{-	0.325	} $ & $	-0.372	^{+	0.031	}_{-	0.038	} $ & $	-0.339	^{+	0.117	}_{-	0.088	} $ & $	-0.351	^{+	0.099	}_{-	0.066	} $ \\
F18, $\sigma_r$	&	1000	& $	-0.417	^{+	0.081	}_{-	0.061	} $ & $	-0.301	^{+	0.032	}_{-	0.060	} $ & $	-0.376	^{+	0.112	}_{-	0.439	} $ & $	-0.177	^{+	0.107	}_{-	0.328	} $ & $	-0.345	^{+	0.025	}_{-	0.047	} $ & $	-0.346	^{+	0.117	}_{-	0.147	} $ & $	-0.346	^{+	0.089	}_{-	0.127	} $ \\
								
\hline																										
	\end{tabular}
    \caption{Resulting slope measurements and their uncertainties for our different methods and data sets. Here, the ''all'' data set refers to all known exoplanets with orbital period and planetary radius estimates and uncertainties and $P<50$ days and $R_P<4.0R_E$, $\sigma_r$ refers to data sets where we accounted for uncertainties in the planetary radii by drawing 1000 planets with replacement from the data set and drawing each planet's radius from a normal distribution centered on the nominal value with standard deviation of the uncertainty in the measurement, and $P<25$ refers to adding a cut to orbital period at 25 days to help mitigate the effects of incompleteness. The other samples refer to data sets used in F17: \citet{Fulton2017}, V18: \citet{vaneylen2018}, and F18: \citet{Fulton2018}. See Section~\ref{sec: data} for a description of each data set. ''DBSCAN'' is density-based clustering, ''HIER'' is hierarchical clustering cut at two branches, ''KDE'' is 2D kernel density estimation, ''CLUST'' is the results from all clustering methods (excluding KDE), and ''ALL M'' is the results from all methods. See Section~\ref{sec: methods} for a description and comparison of all methods. All measurements from all methods and datasets are consistent with a negative slope. We take our bestfit slope to be the result of ''all'' data set and all methods, in bold in the table above, $m=-0.319^{+	0.088	}_{-	0.116	} $. }
    \label{tab:slopes}
\end{table*}


\section{Discussion} \label{sec:disc}

\begin{figure*}
    \centering
    \includegraphics[width=0.48\textwidth]{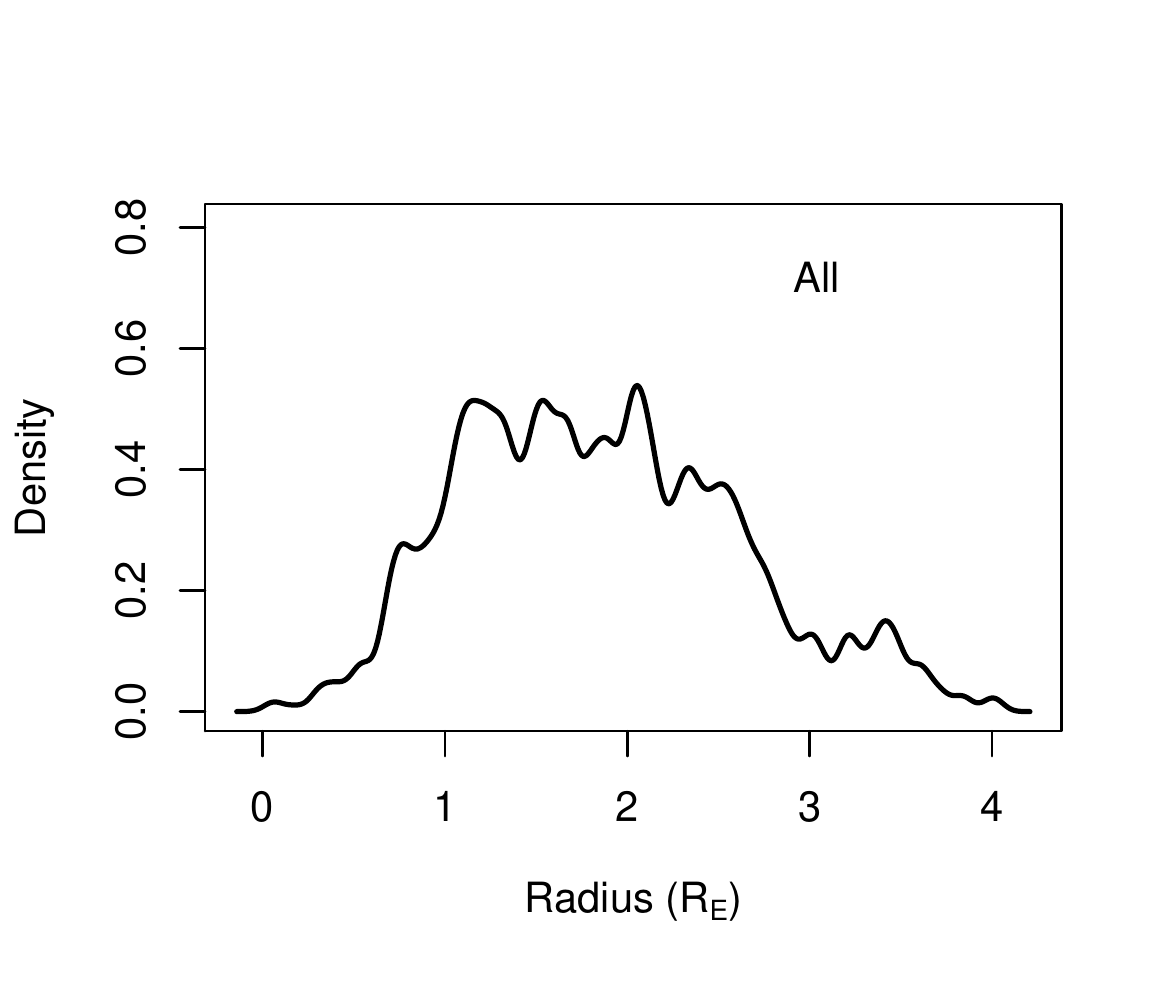}
    \includegraphics[width=0.48\textwidth]{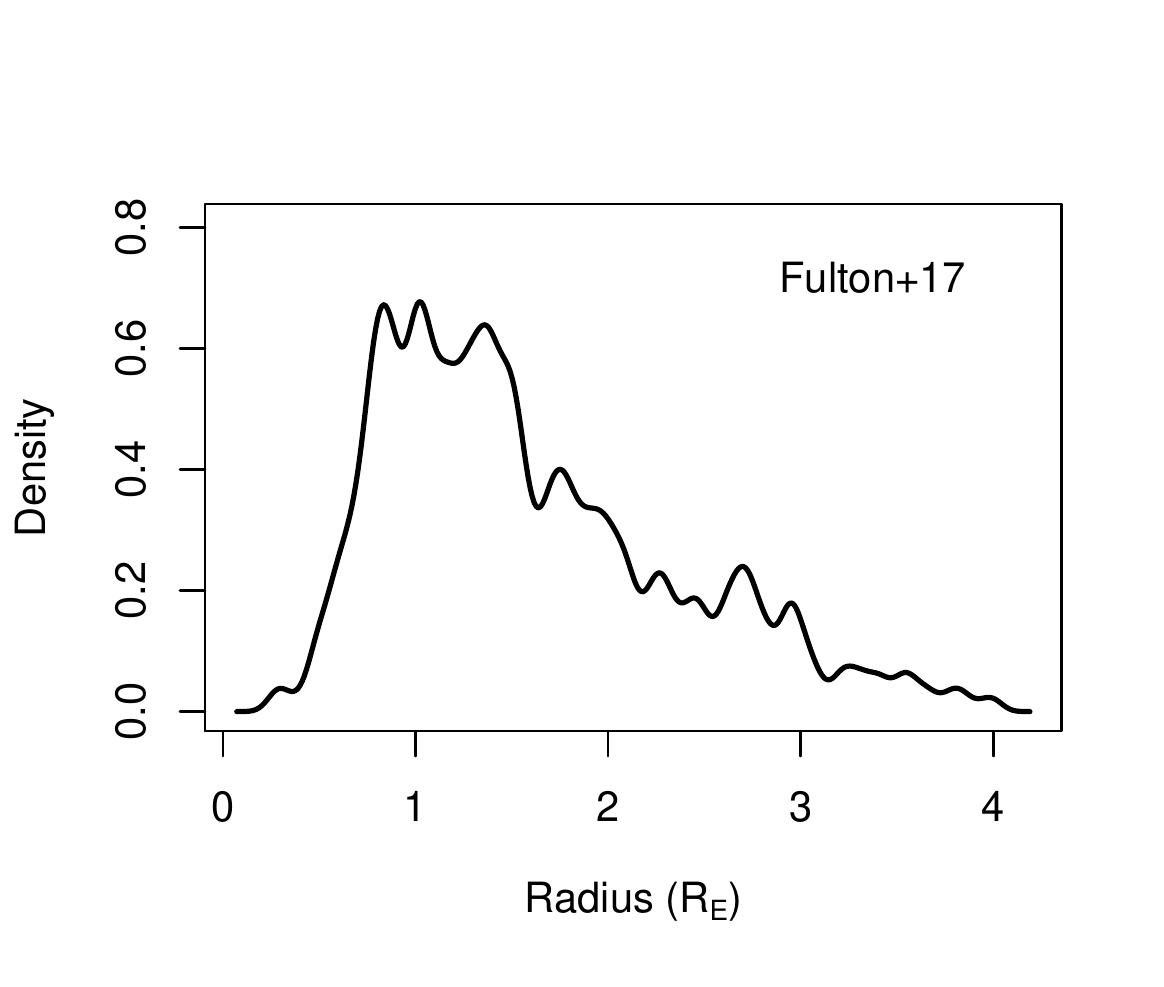}
    \includegraphics[width=0.48\textwidth]{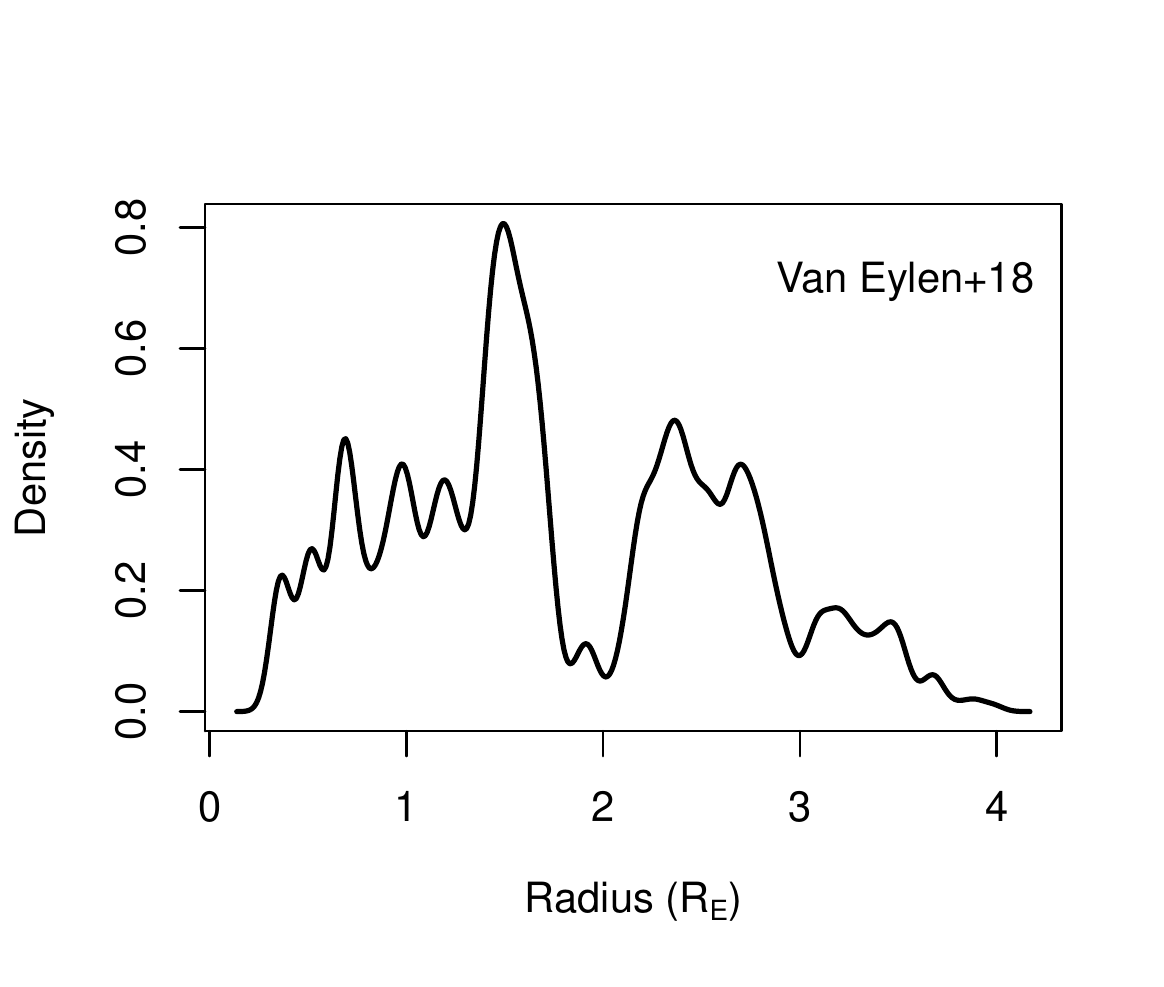}
    \includegraphics[width=0.48\textwidth]{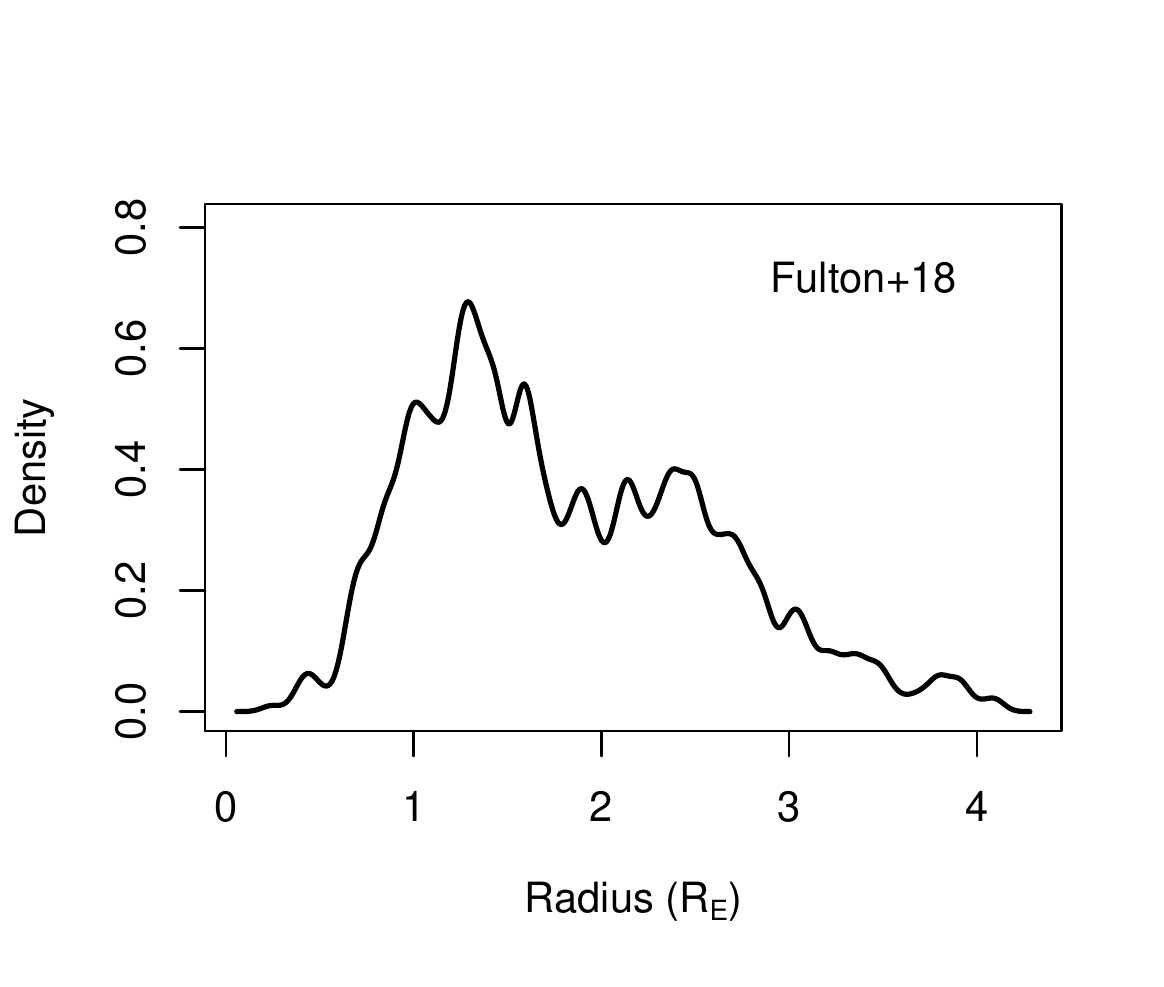}
    \caption{Kernel density estimations of the distributions of planetary radii for our four data sets, where ''all'' is all known exoplanets with orbital period and planetary radius estimates and uncertainties and $P<50$ days and $R_P<4.0R_E$. We accounted for uncertainties in planetary radius and different sizes of the data sets by creating samples of 1000 planets drawn with replacement from each data set, and then drawing a radius for each planet from a normal distribution with the published value as the mean and the uncertainty as the standard deviation. A gap in the distributions of radii was noted by \citet{Fulton2017}, \citet{vaneylen2018}, and \citet{Fulton2018}, but is only distinct in the sample from \citet{vaneylen2018} and noticeable in the sample from \citet{Fulton2018}. A gap is not discernible in the other two data sets.}
    \label{fig:radii}
\end{figure*}

We examine the distributions of planetary radii for each of the data sets in an attempt to observe the gap noted by previous studies \citep[e.g.,][]{Fulton2017,vaneylen2018}. First, we create samples of equal size from each data set by drawing with replacement 1000 times. We also incorporate uncertainties in the planetary radii by drawing each planet radius from a normal distribution that is centered on its nominal value with a standard deviation of the published uncertainty. Our sample of all exoplanets with $P<50$ days and $R_P<4.0$ has a much larger average uncertainty for planetary radii than the other samples (see Table~\ref{tab:data} for a full comparison of our data sets). In Figure~\ref{fig:radii}, we plot a Kernel Density Estimation (KDE) for the radii distributions for each of our re-sampled data sets. We use a KDE to minimize dependence on bin width, although a KDE does require a bandwidth. We note that, once a KDE is used and uncertainties are accounted for, the gap in distribution of planetary radii is barely distinguishable for data set F18, and not at all distinguishable for the F17 and ''all'' data sets. The gap is still prominent in data set V18, which has few planets (81) and the smallest radii uncertainties ($<\sigma_R>=0.06R_E$).

Accounting for the uncertainty in planetary radii obscures not only the gap in the distribution of radii, but also the radius valley as a function of period. When we cluster the ''all'' data set, the slope of the line between the two populations grows from $m=-0.319^{+0.088}_{-0.116}$ using only nominal values for the radii to $m=-0.531^{+0.265}_{-0.401}$ when we account for uncertainties. This large increase in slope is due to the clustering algorithms failing. Instead of clustering along $R_p\sim2R_E$, they tend to cluster along $P\sim7$ days, which results in near vertical lines. This effect becomes even more apparent when we only use planets with $P<25$ days, as can be seen in Table~\ref{tab:slopes}.

We repeat our analysis with a cut in orbital period of $P<25$ days to help mitigate the effects of incompleteness due to observational biases. The resulting slope measurements are consistent with the measurements from the full sample, although larger and less precise (see Table~\ref{tab:slopes} for all slope measurements).

\begin{figure}
    \centering
    \includegraphics[width=0.48\textwidth]{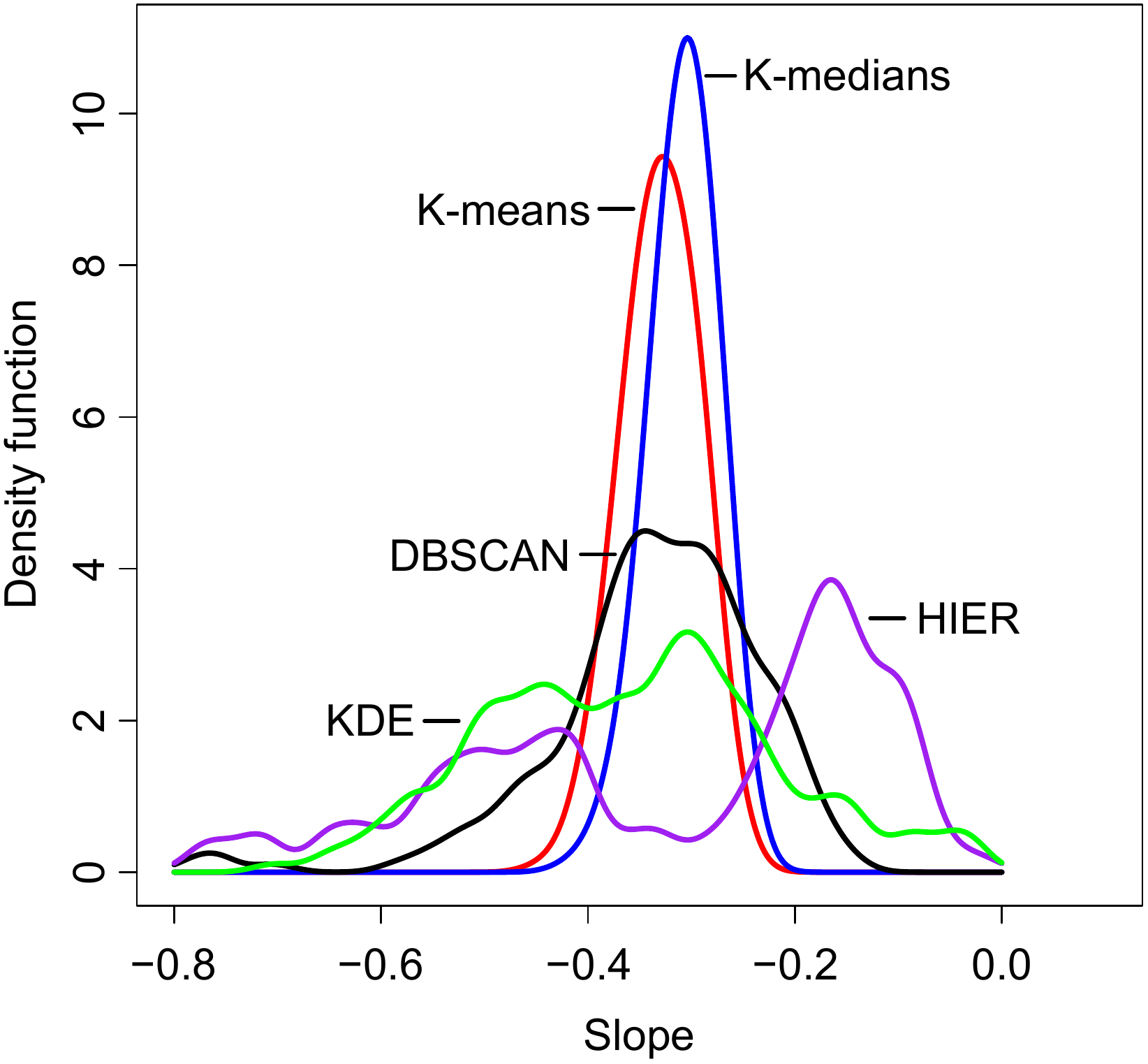}
    \caption{Kernel density estimations of the slopes measured via our different methods for all known exoplanets with orbital period and planetary radius estimates and uncertainties and $P<50$ days and $R_P<4.0R_E$. Here, ''DBSCAN'' is density-based clustering, ''HIER'' is hierarchical clustering (cut at two branches), and ''KDE'' is 2D kernel density estimation (see Section~\ref{sec: methods} for a description of all methods). We take the median and uncertainties from all slopes from all methods as our bestfit slope $m=-0.319^{+0.088}_{-0.116}$.}
    \label{fig:slope}
\end{figure}

The different methods that we employ result in different slopes. In Figure~\ref{fig:slope}, we plot the KDE of the resulting slopes from our analysis on all exoplanets using the nominal values for radius (i.e., not accounting for uncertainties). Slopes larger than $m\sim-0.5$ are a result of the clustering methods failing, which is most apparent in the hierarchical clustering. This method creates an entire tree of classification and we only cut it at two branches. For the lines with steep slopes, a cut at three branches did result in a cluster with $R_p<2.5R_E$ and a cluster with $R_p>2.5R_E$, but we only include the results of the two-branch cut. $K$-means and $K$-medians clustering result in much higher precision than the other methods (for the ''all'' data set as well as most of the other data sets). We take the median and uncertainties from all slopes from all methods as our bestfit slope $m=-0.319^{+0.088}_{-0.116}$.

Our measured slopes are much steeper than those reported in \citet{vaneylen2018} and from theoretical models of photoevaporation and core-powered mass loss, although they are still consistent within $3\sigma$. \citet{vaneylen2018} fit a line to their absence of data and then used support vector machines to determine the hyperplane of maximum separation between the planets above and below the valley, resulting in slope of $m=-0.10\pm0.03$ and $m=-0.09^{+0.02}_{0.04}$, respectively. Given that the other data sets have planets occupying the radius valley, we cannot employ their methods on other data sets. Our methods (discussed in full in Section~\ref{sec: methods}) result in a steeper slope when even we use the smaller asteroseismic data set from \citet{vaneylen2018}. This difference is due to the different classification techniques; as can be seen in comparing our classification (Figure~\ref{fig:clust}) with the classification from support vector machines \citep[Fig. 7 in ][]{vaneylen2018}, the methods employed herein tend to classify planets with periods $P>10$ days as part of the mini-Neptune population.

There is a dearth of mini-Neptunes at short orbital periods ($P<3$ days), that is predicted by photoevaporation models \citep[e.g.,][]{Owen2013} and noted by many authors \citep[e.g.,][]{Lundkvist2018}. This absence of detected planets makes a slope difficult to constrain. Coupled with the observational bias of super-Earths far from their hosts being difficult to detect, this dearth of planets can confuse the clustering and lead to positive slopes. How much this effects the resulting slope is difficult to truly explore, but we do note that the clustering algorithms rarely ($<0.1\%$) result in a positive slope. As mentioned above, we partially investigate this by repeating our analysis with an orbital period cut of $P<25$ days. With this cut, we find slopes that are consistent with the slopes from using all data.

Given that the measured slopes for all data sets and methods are consistent and distinctly negative, we find that the radius valley is consistent with both models of photoevaporation \citep[e.g.,][]{Owen2017} and core-powered mass loss \citep[e.g.,][]{Gupta2018}, but inconsistent with planets forming late in a gas-poor disk \citep[e.g.,][]{Lee2016}. Core-powered mass loss, however, would produce a radius valley that is independent of incident flux, and is therefore independent of stellar type. Recently, \citet{Fulton2018} found that the radius gap shifts to higher incident stellar fluxes around higher mass stars, showing that the gap is dependent on stellar type. This is consistent with predictions from photoevaporation models \citep{Owen2017}. We note that these mechanisms are not mutually exclusive. It is plausible that both photoevaporation and core-powered mass loss, as well as other mechanisms, are responsible for the radius valley simultaneously or at different times during planetary formation.

\begin{figure}
    \centering
    \includegraphics[width=0.48\textwidth]{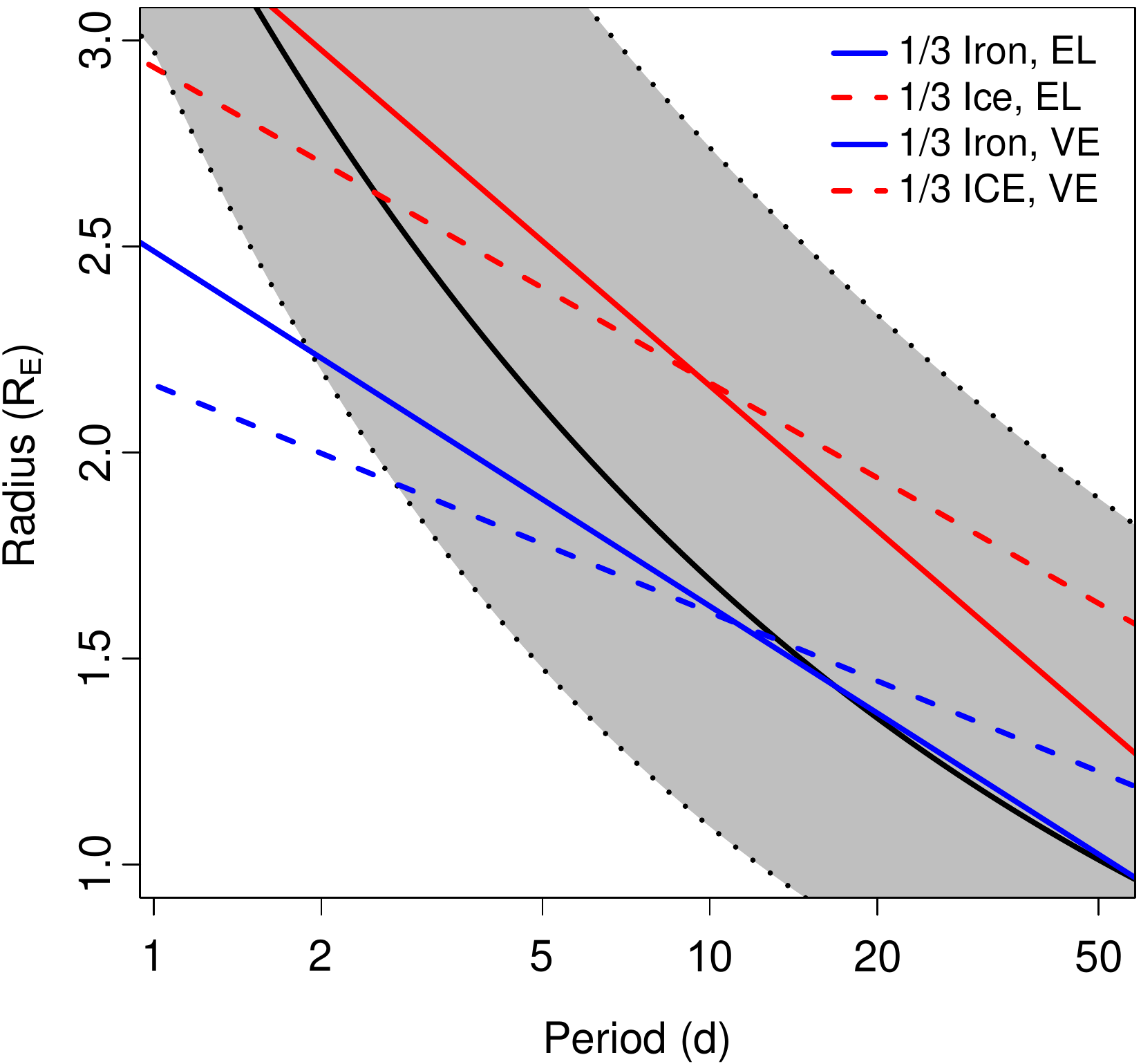}
    \caption{The resulting slope from all methods for all known exoplanets with orbital period and planetary radius estimates and uncertainties and $P<50$ days and $R_P<4.0R_E$ in black and its $1\sigma$ confidence interval in grey ($m=-0.319^{+0.088}_{-0.116}$). We compare the measured slope to theoretical models with different planet-core compositions from \citet{Owen2017} that show the largest super-Earth at each orbital period. Since our slope is the slope of the line separating the two populations and not of the largest super-Earth, it is a bit of an over-estimation. The solid lines are for constant, energy-limited efficiency models (EL) while the dashed lines are for models with variable efficiency \citep[VE, see e.g. ][]{Owen2012}. The blue lines show planets that consist of 1/3 iron and 2/3 silicates while the red lines show planets that consist of 1/3 ice and 2/3 silicates. We find that our measurements are consistent with all four models until an orbital period of $P\sim2.5$ days and are consistent with both models for icy cores for the entire period range. Our measurements are consistent with all four models for the range or orbital periods at $3\sigma$.}
    \label{fig:form}
\end{figure}

The precise slope of the radius valley depends on the planet formation model and the composition of the planets. We compare our resulting slope with different models of photoevaporation from \citet{Owen2017} in Figure~\ref{fig:form}. The photoevaporation models use the maximum radius at the bottom of the valley, but since the observational valley is more of a dearth of planets than a complete absence of planets, we do not have a ''bottom'' of valley to directly compare to. Instead, we plot our resulting slope from all methods for our ''all'' data set with a $1\sigma$ confidence interval, which is an overestimation for what we are comparing to.

Previous studies have found that the location of the radius valley is more consistent with iron-rich cores than with icy cores, assuming that the valley is caused primarily by photoevaporation \citep{Owen2017,jin2018,vaneylen2018}. We find that our measurements are consistent at $1\sigma$ with both the more complex models that include recombination and x-ray evaporation and with the steeper slopes predicted for pure energy-limited evaporation for both iron-rich and icy cores up until an orbital period of $P\sim2.5$ days (although our measurements are consistent with all four models from \citet{Owen2017} at $3\sigma$). Our slopes are more consistent with icy cores than iron-rich cores with orbital periods shorter than $P\sim2.5$ days, possibly indicating a need for including planetary migration after the planets form.


\section{Conclusion}\label{sec: conc}
We investigate the relationship between planetary radius and orbital period, confirming the ``radius valley" that is suggested to separate super-Earths from mini-Neptunes using 2-dimensional kernel density estimator and various clustering techniques. Unlike previous studies, we use a large sample of exoplanets. We use all known exoplanets, but limit our sample to planets with orbital period and planetary radius measurements and $P<50$ days and $R_P<4.0R_E$, resulting in 2066 planets. We repeat our analysis for a sample with planetary radii drawn from normal distributions to account for the uncertainty in the measurements, which can obfuscate the radius valley, and for a sample with $P<25$ days to help mitigate the effects of incompleteness due to observational biases. Accounting for uncertainties and a smaller period cut both led to estimates of the slope that were larger and less precise than using the full sample with nominal values for the radii, but still consistent with each other and our other estimates. We also repeat our analysis on the data sets from \citet{Fulton2017}, \citet{vaneylen2018}, and \citet{Fulton2018}, measuring slopes that are consistent with our other analyses. We present all of our measurements of the radius valley for each method and data set in Table~\ref{tab:slopes}.

\noindent We conclude the following results:
\begin{itemize}
\item Powerful machine-learning techniques allow us to detect and characterize the radius valley using a large sample of exoplanets.
\item The transitional radius between super-Earth and mini-Neptune decreases as a function of orbital period. We find a negative slope which is consistent with models of photoevaporation and core-powered mass loss, but is inconsistent with late formation in a gas-poor disk (characterized by a positive slope).
\item We measure the location of the radius valley as a power law: $\textrm{log}~R_p = m\cdot \textrm{log}~P + b$, where $m=-0.319^{+0.088}_{-0.116}$ and $b=1.26^{+0.28}_{-0.17}$.
\item Given the lack of a clear valley, our sample consists of planets with a wide range of core compositions or planets which have formed beyond the snow line. Our measured slope is consistent at $1\sigma$ with theoretical models of photoevaporation for cores consisting of a significant fraction of iron and for cores consisting of 1/3 ice, 2/3 silicates for planets with $P>2.5$ days. For smaller orbital periods, our measurements are more consistent with icy cores.
\end{itemize}

Given the exoplanetary missions that have recently launched or will soon launch and the missions that have been proposed for the future, such an analysis as presented herein should be repeated after more data have been gathered. The addition of data, especially data that are less biased or have different biases, could grandly expand the significance of this study, providing additional information about the formation and evolution of exoplanetary systems.

\section*{Acknowledgments}
We thank the referee for helpful comments and suggestions that have improved this manuscript. We thank Kathryn Disher for her helpful discussion. This material is based upon work supported by the National Science Foundation Graduate Research Fellowship Program under Grant No. DGE1255832. Any opinions, findings, and conclusions or recommendations expressed in this material are those of the author and do not necessarily reflect the views of the National Science Foundation. This research has made use of the NASA Exoplanet Archive, which is operated by the California Institute of Technology, under contract with the National Aeronautics and Space Administration under the Exoplanet Exploration Program. 

\bibliographystyle{mnras}
\bibliography{all}

\bsp	
\label{lastpage}
\end{document}